\documentclass[12pt]{article}
\usepackage{graphicx}
\usepackage{amsmath}
\usepackage{subfig}
\usepackage{cite}
\usepackage[left=1.5 cm,right= 1.5 cm]{geometry}

\title{\vspace{-4cm}\hspace{15cm}{\small YITP-11-65}\\
\vspace{2cm}Theoretical support for the $\pi (1300)$  and the recently claimed $f_0(1790)$ as molecular resonances}
\author{A. Mart\'inez Torres$^1$, K. P. Khemchandani$^2$,  D. Jido$^1$, A. Hosaka$^2$}
\date{}

\begin{document}

\maketitle

\begin{center}
$^1$ Yukawa Institute for Theoretical Physics, Kyoto University,\\ Kyoto 606-8502, Japan.\\
$^2$Research Center for Nuclear Physics (RCNP), Mihogaoka 10-1,\\ Ibaraki 567-0047, Japan. 
\end{center}

\abstract{A study of three-pseudoscalar  $\pi K \bar{K}$ and $\pi \pi \eta$ coupled system is made by solving the Faddeev equations within an approach based on unitary chiral dynamics. A resonance with total isospin one and spin-parity $J^\pi  = 0^-$ is found with mass $\sim$ 1400 MeV when the $K \bar{K}$ system gets reorganized as the $f_0(980)$.  This resonance is identified with the $\pi (1300)$ listed by the Particle Data Group. Further, the two-body amplitude which describes the interaction between a $\pi$ and the $f_0(980)$ is extracted from the study of the $\pi K\bar K$ and $\pi\pi\eta$ system and is then employed to study the $f_0(980)\pi \pi $ system. As a result, 
a scalar resonance is found near 1790 MeV which drives the two $ f_0 (980)\pi$ systems to resonate as the $\pi (1300)$ while the invariant mass of the two pions falls in the mass region of the scalar $\sigma (600)$. These findings support the existence of a new $f_0$ resonance near 1790 MeV, as found by the  BES  and Crystal Barrel collaborations. Our results
show that this $f_0(1790)$ is definitely distinct to  $f_0(1710)$, the latter of which seems to  possess a glueball structure dominantly.}

\vspace{0.5cm}

\section{Introduction}
The study of light mesons and its spectroscopy is very important to elucidate the working of the strong interactions
at low energies. However, it is a very cumbersome task to understand the properties of these
systems since very different structures with same quantum numbers, like glueballs, hybrids, multi-quark states, molecular states, etc., can exist simultaneously.  
One example is the low-lying scalars meson sector, whose identification is a long-standing puzzle and whose structure is still being debated (see  \cite{scalar1,scalar2,scalar3,scalar4,scalar6,scalar10,scalar11,scalar13,scalar14,kojo,scalar16} as some examples or alternatively see the reviews in the Particle Data Book (PDB) \cite{pdg}).

For the scalar resonances above 1 GeV the situation is also fuzzy, and various configurations like $q\bar q$ states, $q\bar q$ states folded with meson-meson or four quark components, $q\bar q$ components mixed up with glueballs, pure glueballs states, $\rho\rho$ and $K^*\bar K^*$  molecules, etc., have been proposed to explain the nature and properties of resonances like $f_0(1370)$, $f_0(1500)$, $f_0(1710)$ \cite{Fari, Ume, Mauro, Close, Zhao, Klee, Ani, Albadalejo, Raquel, Geng}.  Out of these, the latter two have been widely proposed as the lightest scalar  glueballs (some favoring $f_0(1500)$ and others $f_0(1710)$). To add to the confusion, recently, a new $f_0$ state around 1790 MeV has been found \cite{Bes1, Bes2,cb,Bugg} which is claimed to differ from the $f_0(1710)$ in: i) the signal of the $f_0(1710)$ in the $\omega K\bar K$ and $\phi K\bar K$ data is very clear \cite{Bes1}, while the $f_0(1790)$ is not visible and ii) there is a peak in the $\phi\pi\pi$ data corresponding to the $f_0(1790)$, while the signal for the $f_0(1710)$ is negligible. As a consequence it turns out that the decay of the $f_0(1790)$ into two pions is strong, but there is little or no corresponding signal for its decay to $K\bar K$ (the branching ratios between $K\bar K$ and $\pi\pi$ are larger than a factor of 20). This behavior is not compatible with the properties of the $f_0(1710)$, which is known to decay dominantly to $K\bar K$. These facts, according to Ref.~\cite{Bes1}, indicate the presence of two different states in the data, the $f_0(1710)$ and the $f_0(1790)$.

The panorama does not differ much for the pseudoscalar meson sector, where rather scarce information is available. For example, the first known radial excited state of the pion ($1^-\left( 0^{-+} \right)$) exists with a mass of about 1300 MeV \cite{pdg}, which is  practically ten times greater than the one of its ground state.  Beyond this,  only one more known $1^-\left( 0^{-+} \right)$  resonance is listed in the PDB \cite{pdg}, near 1800 MeV, and there is some debate going on to understand the decay channels of this state within a quarkonia picture  \cite{Barnes,Klempt}. In the isospin zero sector, the nature and properties of the states $\eta(1295)$, $\eta(1405)$ and $\eta(1475)$ have been subject of numerous studies, theoretical as well as experimental, and they are still controversial \cite{pdg}. Recently, an attempt  to shed some light in this area has been made by  studying two-body systems made of a pseudoscalar meson and the resonances $f_0(980)$ or $a_0(980)$ \cite{Albadalejo2} and states which can be associated with the $K(1460)$, $\pi (1300)$, $\pi(1800)$, $\eta(1475)$ and $X(1835)$ have been found using an extension of the model developed in Ref.~\cite{Luis}. Although, some remarks concerning the model used in Ref.~\cite{Luis} have been made in Ref.~\cite{Javi}.  The $K(1460)$ state has been also found as a dynamically generated resonance in the $KK\bar K$ system and coupled channels with the $f_0(980)$ formed in the $K\bar K$ subsystem \cite{AJ3}.

In the present article we discuss a study  which we started by considering  three  pseudoscalars with total strangeness zero. Our approach \cite{mko1,mko2,mko3,mko4,AJ2} involves solving coupled channel Bethe-Salpeter equations within a unitarized approach for the two-hadron subsystems, with the leading order terms obtained using chiral Lagrangians \cite{scalar2,scalar3,chpt3,osetramos,dgr6,dgr7,dgr8}. The resulting two-body amplitudes reproduce the corresponding data and relevant known resonances. Such amplitudes are then used as an input to solve coupled channel Faddeev equations. As we shall show and discuss in detail in this article, we find a pion resonance in our work which can be identified with the $\pi (1300)$ listed in the PDB \cite{pdg}. We find that when the resonance is developed in the $\pi K \bar{K}$ system the $K \bar{K}$ pair gets assembled as the $f_0 (980)$. We obtain the
$ f_0(980)\pi$ amplitude from these results and study yet another three-meson coupled system:  $f_0(980) \pi \pi$, $f_0(980) K \bar{K}$ and find a scalar resonance with mass close to 1790 MeV with the constituting two $ f_0(980)\pi$ systems forming the $\pi (1300)$  and the invariant mass of the two pseudoscalars being in the region of the $\sigma$ resonance.  We discuss that this state is different from the $f_0(1710)$ and its properties are very similar to the state recently found by the
BES and Crystal Barrel collaborations \cite{Bes1,Bes2,cb}, hence, supporting the proposal of existence of a new $f_0$ state with mass 1790 and with properties different to those of  the known $f_0(1710)$ \cite{pdg}.

\section{Formulation}
The formalism used in this manuscript to study three-body systems has been  explained elaborately in Refs.~\cite{mko1,mko2,mko3,mko4,AJ2} and
we refer the reader to these papers for details. Here we limit ourselves to summarizing the main ingredients of the formalism.

\subsection{Two-body and three-body amplitudes}
The starting point in our approach is the determination of the two-body $t$-matrices which describe the interaction between
the pairs present in the system. Once we have obtained these $t$-matrices, we can use them as inputs to solve
the Faddeev equations and determine the three-body $T$-matrix of the system. The calculation of the two-body $t$-matrices is based on solving the Bethe-Salpeter equation implementing unitarity in coupled channels and using the on-shell factorization form of the $t$-matrix \cite{scalar2,scalar3}, 
\begin{equation}
t=(1-V\tilde{G})^{-1}V. \label{BS}
\end{equation}
In this method, the kernel $V$ in Eq.~(\ref{BS})  corresponds to the lowest order two-body amplitude obtained from chiral Lagrangians \cite{scalar2} and $\tilde{G}$
represents the loop function of two hadrons. In the present case, i.e., for a two pseudoscalar system, we calculate this loop function using the dimensional regularization scheme of Ref.~\cite{scalar3},

\begin{align}
\tilde{G}_r=&\frac{1}{16\pi^2}\Bigg\{a_r(\mu)+\ln\frac{m^2_{1r}}{\mu^2}+\frac{m^2_{2r}-m^2_{1r}+E^2}{2E^2}\ln\frac{m^2_{2r}}{m^2_{1r}}\nonumber\\
&+\frac{q_r}{E}\Bigg[\ln\Big(E^2-(m^2_{1r}-m^2_{2r})+2q_rE\Big)+\ln\Big(E^2+(m^2_{1r}-m^2_{2r})+2q_rE\Big)\nonumber\\
&-\ln\Big(-E^2+(m^2_{1r}-m^2_{2r})+2q_rE\Big)-\ln\Big(-E^2-(m^2_{1r}-m^2_{2r})+2q_rE\Big)\Bigg]\Bigg\}.\label{g}
\end{align}
In Eq.~(\ref{g}), $E$ is the total energy of the two-body system, $m_{1r}$, $m_{2r}$ and $q_r$ correspond, respectively, to the masses and the center of mass momentum of the two pseudoscalars present in the r{\it th} channel, $\mu$ is a regularization scale and $a_r(\mu)$ a subtraction constant. Following  Ref.~\cite{scalar3}, we have taken $\mu =1224$ MeV  and a
value for $a_r(\mu)\sim-1$. In this way we can reproduce the observed two-body phase shifts and inelasticities for the different  coupled channels as done in Refs.~\cite{scalar2,scalar3}. 

After evaluating the two-body amplitudes, we can proceed with the determination of the three-body $T$ matrix for the system  considered. To do that we use the approach of Refs.~\cite{mko1,mko2,mko3,mko4,AJ2}, in which the Faddeev partitions, $T^1$, $T^2$ and $T^3$,  are written as 
\begin{equation}
T^i =t^i\delta^3(\vec{k}^{\,\prime}_i-\vec{k}_i) + \sum_{j\neq i=1}^3T_R^{ij}, \quad i=1,2,3,
\label{Ti}
\end{equation}
with $\vec{k}_{i}$ ($\vec{k}^\prime_{i}$) being the initial (final) momentum of the particle $i$ and $t^{i}$ the two-body $t$-matrix which describes the interaction of the $(jk)$ pair of the system, $j \neq k\neq i=1,2,3$. Using Eq.~(\ref{Ti}), the full three-body $T$-matrix is obtained in terms of the two-body $t$-matrices and the $T^{ij}_{R}$  partitions as
\begin{equation}
T = T^{1} + T^{2} + T^{3} = \sum_{i=1}^{3}t^i\delta^3(\vec{k}^{\,\prime}_i-\vec{k}_i) +T_{R}\label{T}
\end{equation}
where we define
\begin{equation}
T_{R} \equiv \sum_{i=1}^3\sum_{j\neq i=1}^{3}T^{ij}_{R} . \label{ourfullt}
\end{equation}

The $T^{ij}_{R}$ partitions in Eq.~(\ref{Ti}) satisfy the following set of coupled equations

\begin{equation}
T^{\,ij}_R = t^ig^{ij}t^j+t^i\Big[G^{\,iji\,}T^{\,ji}_R+G^{\,ijk\,}T^{\,jk}_R\Big], \quad i\ne j, j\ne k = 1,2,3. 
  \label{Trest}
\end{equation}
where $g^{ij}$ corresponds to the three-body Green's function of the system and its elements are defined as
\begin{eqnarray}
g^{ij} (\vec{k}^\prime_i, \vec{k}_j)=\Bigg(\frac{N_{k}}{2E_k(\vec{k}^\prime_i+\vec{k}_j)}\Bigg)\frac{1}{\sqrt{s}-E_i
(\vec{k}^\prime_i)-E_j(\vec{k}_j)-E_k(\vec{k}^\prime_i+\vec{k}_j)+i\epsilon},\label{Green}
\end{eqnarray}
with $N_{k}=1$ for mesons  and $E_{l}$, $l=1,2,3$, is the energy of the particle $l$.

The $G^{ijk}$ matrix in Eq.~(\ref{Trest}) represents  a loop function of three-particles and it is written as
\begin{equation}
G^{i\,j\,k}  =\int\frac{d^3 k^{\prime\prime}}{(2\pi)^3}\tilde{g}^{ij} \cdot F^{i\,j \,k}
\label{eq:Gfunc}
\end{equation}
with the elements of  $\tilde{g}^{ij}$ being 
\begin{eqnarray}
\tilde{g}^{ij} (\vec{k}^{\prime \prime}, s_{lm}) = \frac{N_l}
{2E_l(\vec{k}^{\prime\prime})} \frac{N_m}{2E_m(\vec{k}^{\prime\prime})} 
\frac{1}{\sqrt{s_{lm}}-E_l(\vec{k}^{\prime\prime})-E_m(\vec{k}^{\prime\prime})
+i\epsilon}, \quad i \ne l \ne m,
\label{eq:G} 
\end{eqnarray}
and the matrix $F^{i\,j\,k}$, with explicit variable dependence, is given by 
\begin{eqnarray}
F^{i\,j\,k} (\vec{k}^{\prime \prime},\vec{k}^\prime_j, \vec{k}_k,  s^{k^{\prime\prime}}_{ru})= t^{j}(s^{k^{\prime\prime}}_{ru}) g^{jk}(\vec{k}^{\prime\prime}, \vec{k}_k)
\Big[g^{jk}(\vec{k}^\prime_j, \vec{k}_k) \Big]^{-1}
\Big[ t^{j} (s_{ru}) \Big]^{-1},  \quad j\ne r\ne u=1,2,3.\label{offac}
\end{eqnarray}
In Eq. (\ref{eq:G}), $\sqrt{s_{lm}}$ is the invariant mass of the $(lm)$ pair and can be calculated in terms of the external variables. The upper index $k^{\prime\prime}$ for the invariant mass $s^{k^{\prime\prime}}_{ru}$ of Eq.~(\ref{offac}) indicates its dependence on the loop variable (see Ref. \cite{mko2} for more details). 

Equations (\ref{Trest}) are based on the finding  of  an analytical cancellation between the contribution of the off-shell parts of the chiral two-body $t$-matrices to the three-body Faddeev diagrams and a contact term with the same topology and whose origin is in the chiral Lagrangian used to describe the interaction (see Refs. \cite{AJ3,mko1,mko2,mko3,mko4,AJ2} for more details). Such a cancellation was shown explicitly for a three pseudoscalar system (with which we are concerned in the present work) to be exact in the chiral limit \cite{AJ3}. In a realistic case,  the sum of the different off-shell sources together with the three-body contact term was estimated to be around $7\%$ of the total on-shell contribution~\cite{AJ3}. Therefore, it has been shown that only the contribution from the on-shell part of the two-body chiral $t$-matrices is significant in our formalism and, hence, we rely on it. 

Finally, the $T^{ij}_R$ partitions given in Eq.~(\ref{Trest}) are calculated as function of  the total three-body energy, $\sqrt s$, and the 
invariant mass of  the particles 2 and 3, $\sqrt{s_{23}}$, and they are determined for real values of these variables. The other invariant masses, $\sqrt{s_{12}}$ and $\sqrt{s_{31}}$ can be obtained in terms of  $\sqrt s$ and $\sqrt{s_{23}}$, as it was shown in Ref. \cite{mko2,mko3}. Peaks  found in the modulus squared  of the three-body $T$-matrix can be associated to dynamically generated resonances. Finally, the first term in Eq.~(\ref{T}) can not give rise to any three-body structure, thus,
to identify possible three-body states we study the properties of the $T_R$ matrix defined in
Eq.~(\ref{ourfullt}).

\begin{figure}[h!]
\centering
\includegraphics[width=0.55\textwidth]{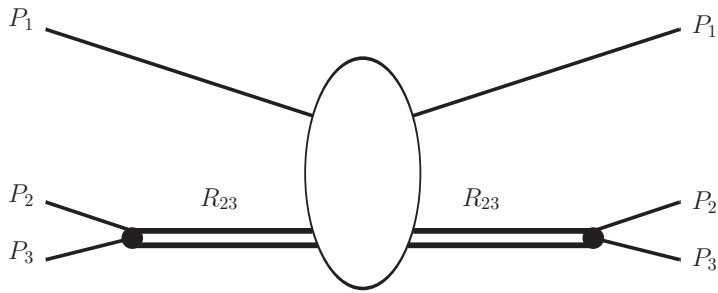}
\caption{Diagrammatic representation of the process $P_1 (P_2 P_3)_{I_{23}}\to P_1 (P_2 P_3)_{I_{23}}$ in terms of the $P_1 R_{23}\to P_1 R_{23}$ amplitude. Here $P_1$, $P_2$, $P_3$ denote particles 1, 2, 3, respectively, and $R_{23}$ represents a resonance formed by the interaction of particles 2 and 3 (see the text for more details).}\label{rel}
\end{figure}

\subsection{Extraction of an effective two-body interaction from the three-body amplitude}\label{frame}
The states found within this approach can be catalogued as molecular hadronic states formed by the interaction of three particles.
It is interesting to notice that if in such molecular states two out of  three particles form a resonance,  the calculated three-body $T$ matrix  can be used to determine the amplitude between the resonance and the other particle which constitutes the three-body system. This is possible if one knows the coupling of the two-body resonance with the particles which constitute it, as we shall discuss below.  As a consequence, it is possible to treat the three-body amplitude obtained in our approach  as an effective two-body (particle + resonance) amplitude. We could then add one more particle to such an effective two-body system and eventually study the dynamics of a four particle system (particle+particle+resonance) by solving three-body equations. 

Let us call  the particles which constitute the three-body system under consideration as $P_1$, $P_2$ and $P_3$ and let us assume that particles $P_2$ and $P_3$ form a resonance, which we denote by $R_{23}$, for a certain isospin $I_{23}$. Following Refs.~\cite{jh,sekihara}, if $g_{R_{23}\to (P_2P_3)}$ is the coupling of the particles $P_2$ and $P_3$ to the resonance $R_{23}$ for the isospin  $I_{23}$,  the amplitude of the three-body system $P_1 P_2 P_3$ can be expressed in terms of the amplitude of the particle+resonance system, i.e., $P_1$+$R_{23}$, as (see Fig.~\ref{rel})

\begin{align}
T_{P_1 (P_2P_3)}(\sqrt{s},\sqrt{s_{23}}\simeq M_{R_{23}})&=g_{R_{23}\to (P_2P_3)}G_{R_{23}}\,t_{P_1 R_{23}}(\sqrt{s})\nonumber\\
&\quad\times G_{R_{23}}\,g_{R_{23}\to (P_2P_3)}.\label{TP1P2P3}
\end{align}
where $G_{R_{23}}$ represents the propagator of the resonance $R_{23}$, and is given by
\begin{equation}
G_{R_{23}}=\frac{1}{s_{23}-M^2_{R_{23}}+i M_{R_{23}}\Gamma_{R_{23}}},\label{GP23}
\end{equation}
with  $M_{R_{23}}$, $\Gamma_{R_{23}}$ being the mass and width, respectively, of the $R_{23}$ resonance generated and $\sqrt{s}$ the total energy of the $P_1 R_{23}$ system. 
Close to the energy region in which the $R_{23}$ resonance is formed, the $P_2P_3\to P_2P_3$ amplitude in isospin $I_{23}$ can be expressed as
\begin{equation}
t_{(P_2P_3)}(\sqrt{s_{23}})=\frac{g^2_{R_{23}\to (P_2P_3)}}{s_{23}-M^2_{R_{23}}+iM_{R_{23}}\Gamma_{R_{23}}}.\label{gP23toP2P3}
\end{equation}
In this way, at the value $\sqrt{s_{23}}= M_{R_{23}}$, using Eqs.~(\ref{GP23})-(\ref{gP23toP2P3}), we can write Eq.~(\ref{TP1P2P3}) like
\begin{equation}
T_{P_1 (P_2P_3)}(\sqrt{s},\sqrt{s_{23}}=M_{R_{23}})=\frac{t_{(P_2P_3)}(\sqrt{s_{23}}=M_{R_{23}})}{iM_{R_{23}}\Gamma_{R_{23}}}t_{P_1 R_{23}}(\sqrt{s}).
\end{equation}
Therefore, the amplitude which describes the $P_1 R_{23}$ interaction can be calculated as

\begin{equation}
t_{P_1 R_{23}}(\sqrt{s})=\frac{iM_{R_{23}}\Gamma_{R_{23}}}{t_{(P_2P_3)}(\sqrt{s_{23}}=M_{R_{23}})}T_{P_1 (P_2P_3)}(\sqrt{s},\sqrt{s_{23}}=M_{R_{23}})\label{tP1P23}
\end{equation}
where the two-body and three-body amplitudes $t_{(P_2P_3)}$ and $T_{P_1 (P_2P_3)}$ are obtained by solving Eq.~(\ref{BS}) and Eq.~(\ref{Trest}), respectively, and the mass and width of the $R_{23}$ resonance can be extracted from the pole position of the  two-body amplitude $t_{(P_2P_3)}$ on the complex energy plane.

The t-matrix in Eq.~(\ref{tP1P23}) can be used now as an input for Eq.~(\ref{Trest}) when studying  the interaction between a particle, $P_0$, with the system made of the particle $P_1$ and the resonance $R_{23}$, i.e., considering the system $P_0-P_1-R_{23}$ as an effective three-body system.  In this case, to determine the energy and momenta which appear in Eq.~(\ref{Green}) and Eq.~(\ref{eq:Gfunc}) we use the resonance mass $M_{R_{23}}$.

\section{Results}
\subsection{The $\pi K\bar K$-$\pi\pi\eta$ coupled system. }
Our aim is to look for the possible generation of molecular resonances in the $\pi K\bar K$ and $\pi\pi\eta$ coupled system considering $S$-wave interactions among the constituent pairs. In such a case, a reliable calculation requires the generation  of the S-wave $\sigma(600)$, $f_0(980)$, $a_0(980)$ and $\kappa(850)$ states in the different two-body subsystems, in others words, we need to solve Eq.~(\ref{BS}) within a coupled channel approach. In particular, we take into account all possible two-body channels of two pseudoscalar mesons ($\pi$, $\eta$, $K$, $\bar K$)  which couple to $ K\bar K$, $K\pi $, $K\eta$,  except for the $\eta\eta$ channel, whose effect in the $K \bar K$ amplitude is known to be negligible \cite{scalar2}.  In this way, the isoscalar $K\bar{K}$ and $\pi\pi$ $t$-matrices obtained within this model in S-wave and isospin 0 dynamically generate the resonances $f_0(980)$ and $\sigma(600)$, while the system composed by the channels $K\bar K$ and $\pi\eta$ in isospin 1 gives rise to the $a_0(980)$ state. In the strangeness +1 $K\pi$ and $K\eta$ systems the $\kappa(850)$ is formed.

For the three-body calculations we work with total charge zero, having then seven coupled channels to handle: $\pi^0 K^+ K^-$, $\pi^0 K^0\bar K^0$, $\pi^0\pi^0\eta$, $\pi^+ K^0K^-$, $\pi^+\pi^-\eta$, $\pi^-K^+\bar K^0$, $\pi^-\pi^+\eta$.
All the matrices present in  Eq.~(\ref{Trest}) are projected on S-wave, which implies that the quantum numbers of the three-body system and, thus, the possible bound states or resonances present in it are $J^{\pi}=0^{-}$.  

To identify the peaks obtained in the three-body $T$-matrix for the different channels  with physical states we need to project these amplitudes, which are calculated in the charge basis, on an isospin basis. To do that, we consider the total isospin  $I$ of the three-body system and the isospin of one of the two-body subsystems, which in the present case is taken as the isospin of the $K\bar K$ subsystem or (23) subsystem, $I_{23}$, and evaluate the transition amplitude $\langle I,I_{23}|T_R|I,I_{23}\rangle$. The isospin  $I_{23}$ can be 0 or 1, thus, the total isospin $I$ can be 0, 1 or 2. For the cases involving the states $|I=0$, $I_{23}=1\rangle$,  $|I=1$, $I_{23}=1\rangle$ and $|I=2$, $I_{23}=1\rangle$ we have not found any clear signal which could be related to a resonance or bound state. Thus, in the following, we discuss the case  $I=1$ with $I_{23}=0$, where we find a resonance.

\begin{figure}[h!]
\centering
\includegraphics[width=0.85\textwidth]{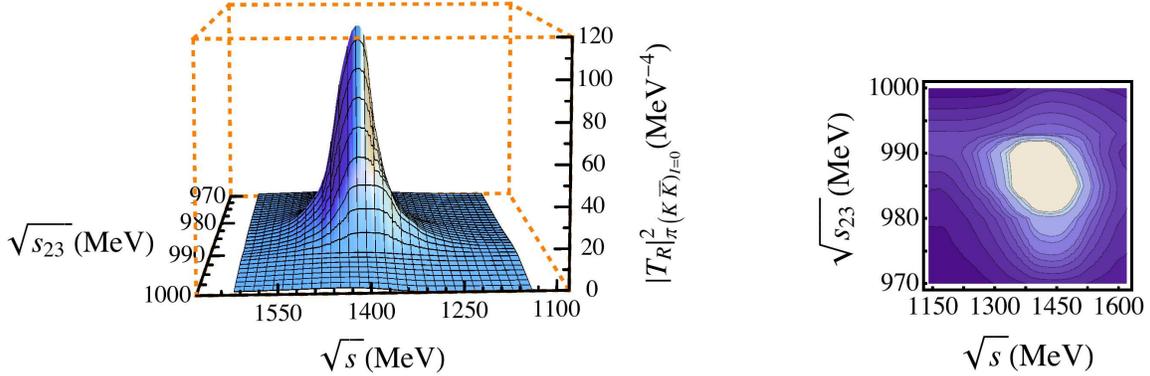}
\caption{(Left) Squared amplitude for the $\pi K\bar K$ channel for total isospin $I=1$ with the $K\bar K$ subsystem in isospin zero. (Right) Contour plot as a function of the total energy, $\sqrt{s}$, and the invariant mass $\sqrt{s_{23}}$ of the $K\bar K$ subsystem, which is in isospin zero.}\label{pif0}
\end{figure}

The modulus squared $\pi K\bar K$ amplitude projected on total isospin 1, with the $K\bar K$ subsystem
in isospin zero, is shown in Fig.~\ref{pif0}.  As one can see, a peak around 1400 MeV with 85 MeV width appears with the invariant mass of the $K\bar K$ subsystem, which is in isospin zero, being around 985 MeV. This means that the $f_0(980)$ state gets dynamically generated in the $K\bar K$ subsystem when the three-body state is formed. Instead of using the isospin base $|I,I_{23}\rangle$ we can also employ the base characterized by the total isospin and the isospin of the (12) subsystem,  i.e., $\pi K$ subsystem, $I_{12}$. In this case we find that when the corresponding peak  shows up at 1400 MeV, the invariant mass  $\sqrt{s_{12}}$ is around 900 MeV, thus, close to the energy region in which the $\kappa(850)$ gets dynamically generated. However, the squared amplitude for the case $I=1$, $I_{23}=0$ is two orders of magnitude bigger than the one for $I=1$, $I_{12}=1/2$, thus, the structure of the state  at 1400 MeV is dominantly $\pi f_0(980)$.

This state can be associated with the $\pi(1300)$  listed in the PDB \cite{pdg}, whose mass is in the range $1300\pm 100$ MeV and the width found from the different experiments listed varies between 120 to 700 MeV \cite{pdg}. 
Using these values as a reference, the peak position obtained here is in the experimental upper limit  for this state, while the width is close to the lower experimental value, thus,  our findings are compatible with the known data set. Surely, for a better comparison one needs more experiments which could help in determining the properties of this state with more precision. The decay modes seen for this resonance are $\rho \pi$ and $\pi (\pi\pi)_{Swave}$. The channel  $\pi \pi\pi$ is a three-body 
channel which couples to $\pi K\bar K$ and $\pi\pi\eta$. However the three pion threshold
(around 410 MeV) is far away from the region in which the state is formed, thus, it naturally is not essential in the generation of the $\pi(1300)$. However, the inclusion of channels like  $\pi \pi\pi$ or  $\rho \pi$ could help in increasing the width found for the state within our approach, since there is more phase space for the $\pi(1300)$ to decay to these channels.

Finally, we would like to mention that apart from the $\pi K\bar K$, $\pi\pi\eta$ systems, we have also studied the $\eta K\bar K$, $\eta\pi\pi$ systems in S-wave to search for possible signals of the $\eta$ states listed in the PDB, $\eta(1295)$, $\eta(1405)$ and $\eta(1475)$, but we have not found any clear signal which could be related to any of them.

\begin{figure}
\centering
\includegraphics[width=0.3\textwidth]{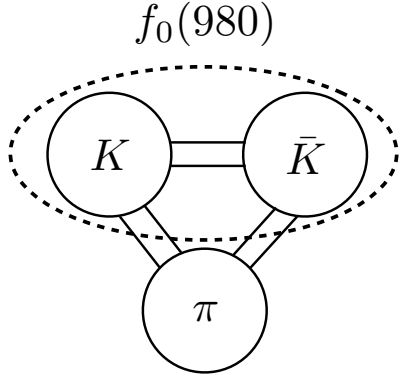}
\caption{Internal structure of the $\pi(1300)$ resonance.}\label{pi_struct}
\end{figure}

\subsection{Study of the $f_0(980)\pi\pi$ and $f_0(980)K\bar K$ systems.}
The state found at 1400 MeV can be understood as a molecular resonance formed by a pion and the $f_0(980)$, which is dynamically generated in the $K\bar K$ interaction (see Fig.~\ref{pi_struct}). As explained in Sec.~\ref{frame}, we can use the obtained $\pi (K\bar K)_{I=0}\to\pi (K\bar K)_{I=0}$ three-body $T$-matrix  to determine the amplitude which describes the interaction between the pion and the $f_0 (980)$, and use this latter one to study the $f_0(980)\pi\pi$ system. To do this, we first need to relate the amplitude of the  $\pi (K\bar K)_{I=0}$ system with the one of the $\pi f_0(980)$ system. For that, particularizing Eq.~(\ref{tP1P23}) for the $\pi$-$f_0(980)$ system, we get

\begin{equation}
t_{\pi f_0(980)}(\sqrt{s_{\pi f_0}})=\frac{iM_{f_0(980)}\Gamma_{f_0(980)}}{t_{(K\bar K)_{I=0}}(\sqrt{s_{23}}=M_{f_0(980)})}T_{\pi (K\bar K)_{I=0}}(\sqrt{s_{\pi f_0}},\sqrt{s_{23}}=M_{f_0(980)})\label{Tpif0}
\end{equation}
where $M_{f_0(980)}=982$ MeV and $\Gamma_{f_0(980)}=32$ MeV represent the mass and width, respectively, of the $f_0(980)$ resonance (these values have been extracted from the $K\bar K$ amplitude calculated on the real plane, which are very similar to the values obtained on the complex plane)  and $\sqrt{s_{\pi f_0}}$ is the total energy of the $\pi$-$f_0(980)$ system.

Similarly, using the result of Ref.~\cite{AJ3}, we can obtain the amplitude for the $K f_0(980)$ system in terms of the amplitude for the $KK\bar K$ system and the one for the $K\bar K$ in isospin 0 like
\begin{equation}
t_{K f_0(980)}(\sqrt{s_{K f_0}})=\frac{iM_{f_0(980)}\Gamma_{f_0(980)}}{t_{(K\bar K)_{I=0}}(\sqrt{s_{23}}=M_{f_0(980)})}T_{K (K\bar K)_{I=0}}(\sqrt{s_{K f_0}},\sqrt{s_{23}}=M_{f_0(980)}).\label{TKf0}
\end{equation}
Using Eq.~(\ref{Tpif0}) and Eq.~(\ref{TKf0}) as inputs for Eq.~(\ref{Trest}) we can study the coupled channel systems $f_0(980)\pi\pi$ and $f_0(980) K\bar K$.

\begin{figure}[h!]
\centering
\hspace{2.5cm}
\includegraphics[width=0.83\textwidth]{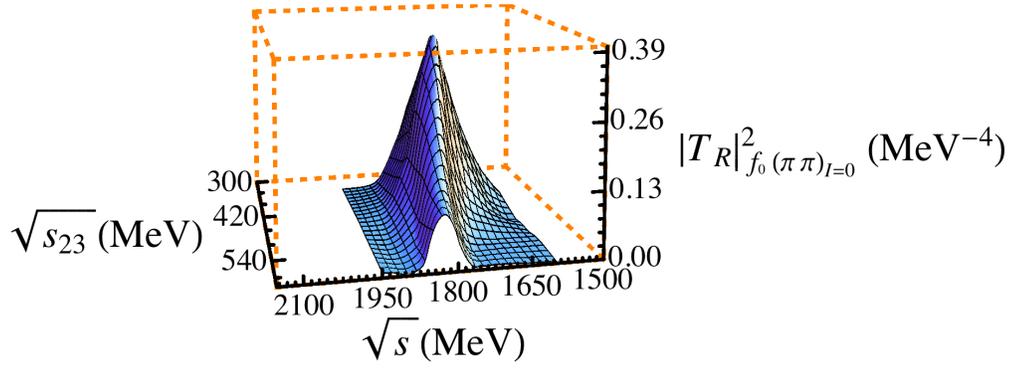}\\
\hspace{-0.91cm}
\includegraphics[width=0.8\textwidth]{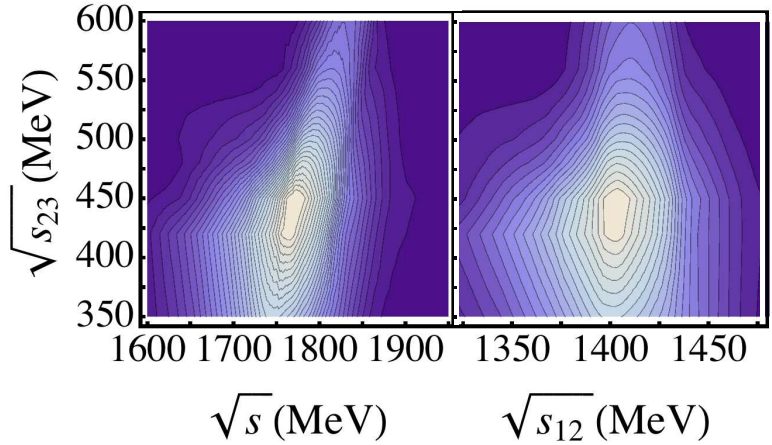}
\caption{(Upper panel) Squared amplitude for the $f_0(980)\pi\pi$ channel for total isospin zero, thus, with the $\pi\pi$ subsystem in isospin zero. (Lower panel) Contour plots as a function of the total energy of the $f_0(980)\pi\pi$ system, $\sqrt{s}$, and the invariant mass of the $\pi\pi$ subsystem, $\sqrt{s_{23}}$ (Left side) and as a function of the 
$\pi\pi$ and $f_0(980)\pi $ invariant masses, $\sqrt{s_{23}}$ and $\sqrt{s_{12}}$, respectively (Right side).}\label{f0pipi}
\end{figure}

In Fig.~\ref{f0pipi} we show the result obtained for the $f_0(980)\pi\pi$ amplitude with the $\pi\pi$ subsystem in isospin zero, thus the three-body system has total quantum numbers $0^{++}$. As one can see, a peak around 1773 MeV with 100 MeV width
develops when  the $\pi\pi$ system is in isospin zero and has an invariant mass around 450 MeV, and the invariant mass of the $f_0(980)\pi$ is around 1400 MeV. The value of the $\pi\pi$ invariant mass at which the peak in the $f_0(980)\pi\pi$ amplitude shows up is in the region of the $\sigma$ resonance, which  gets manifested in the $\pi\pi$ $t$-matrix used to solve Eq.~(\ref{Trest}) at an energy around 600 MeV and with a width around 500 MeV (which are the values associated with the $\pi\pi$ amplitude on the real axis).  In that sense, the generation of the state at 1773 MeV  organizes the $f_0(980)\pi$ subsystems as the $\pi(1300)$ and, although the $\pi\pi$ subsystem is not resonating exactly at the $\sigma$ resonance position, the attraction present in it is important to generate the three-body state. For the resonance found at 1773 MeV, the effect of the $f_0(980) K\bar K$ channel is found to be negligible and one can solve Eq.~(\ref{Trest}) considering only  the channel $f_0(980)\pi\pi$. For higher values of the invariant mass of the $\pi\pi$ and $K\bar K$ subsystems, around the region of the $f_0(980)$ and $a_0(980)$, we do not find a clear structure
which could be associated with higher scalar resonances, like $f_0(2000)$ or $f_0(2100)$.

\begin{figure}[h!]
\centering
\includegraphics[width=0.4\textwidth]{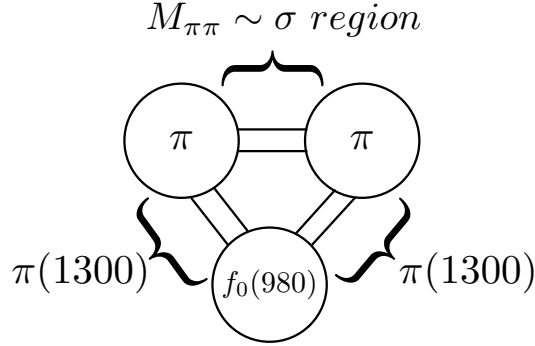}
\caption{Internal structure of the $f_0(1790)$.}\label{f0_struct}
\end{figure}

\begin{figure}[h!]
\centering
\includegraphics[width=0.75\textwidth]{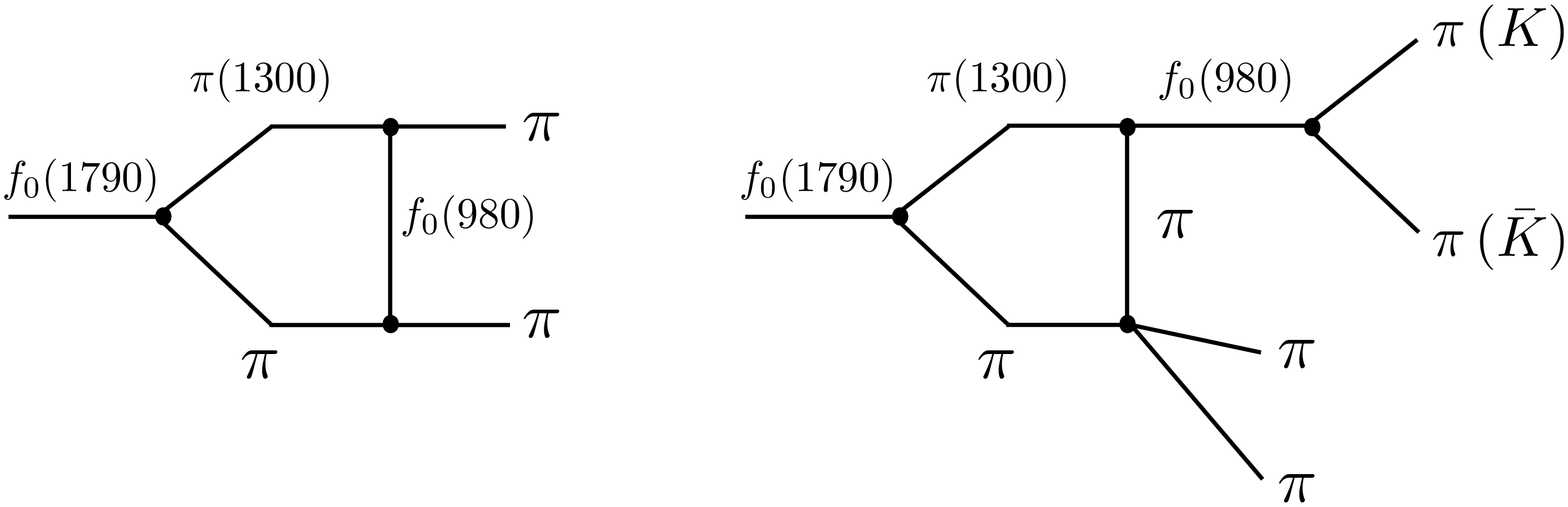}
\caption{Decay modes of  the $f_0(1790)$ found in this work.}\label{decays}
\end{figure}

The scalar resonance found at 1773 MeV can be interpreted as a molecular state of  $\pi$-$\pi(1300)$, with  $\pi(1300)$ being a $\pi f_0(980)$ molecular resonance (see Fig.~\ref{f0_struct}). A state with this structure will decay dominantly to $\pi\pi$, $\pi\pi\pi\pi$ and $\pi\pi K\bar K$, as shown in Fig.~\ref{decays}, having larger phase space for the $\pi\pi$ channel. If we try to associate this resonance with one of the scalar states listed in the PDB, there is only one possibility: the $f_0(1710)$. But this state is known to decay dominantly to $K\bar K$ and  its decay to pions is suppressed \cite{pdg}. This fact is in contradiction with the properties of the scalar resonance found in the present work, which can not decay to $K\bar K$ (as is clear from Fig.~\ref{decays}).    However, a new $f_0$ with mass around 1790 MeV  has been found in two pion spectrum by the BES collaboration in Ref.~\cite{Bes1} and it has also been indicated in an analysis \cite{Bugg} of the 4$\pi$ data from the BES collaboration \cite{Bes2}.  One peculiarity of the $f_0(1790)$ observed in Refs.~\cite{Bes1,Bugg,Bes2} is that its decay to $K\bar K$ is strongly suppressed as compare to its decay to $\pi\pi$ or $\pi\pi\pi\pi$, which is strikingly similar to the characteristics of the $f_0$ resonance found in our present work. Thus, we associate the scalar resonance found  at 1773 MeV with the $f_0(1790)$ found in Refs.~\cite{Bes1,Bugg,Bes2}.

One comment is here in order. The resolution of Eq.~(\ref{Trest}) for the $f_0(980)\pi\pi$ system implies consideration of
intermediate states in which two pions and a $f_0(980)$ are propagating. This does not have to be necessarily true always, since we could have enough energy to excite the $f_0(980)$ resonance and have then intermediate states of four particles, i.e., $\pi\pi K\bar K$. However, the fact that a signal is observed around 1773 MeV when we rely only on the propagation of the $f_0(980)$ resonance and not of a $K$ and a $\bar K$  indicates that the consideration of $\pi\pi K\bar K$ intermediate states  contributes mainly to the background, which basically could produce an increase in the width of the state.

\section{Conclusions}
We have investigated the $\pi K\bar K$ system and coupled channels in S-wave using an approach based on solving the Faddeev equations within the use of unitary chiral dynamics to determine the two-body input $t$-matrices. The study has revealed the formation of a state within isospin 1, $J^\pi=0^-$, mass around 1400 MeV and width of 85 MeV which can be associated with the $\pi(1300)$ listed in the Particle Data Book. The  generation of this state configures the $K\bar K$ subsystem as the $f_0(980)$ resonance. Later on, considering the state found at 1400 MeV as an effective $\pi$-$f_0(980)$ system, we have related the $\pi K\bar K$ and $\pi f_0(980)$ amplitudes using the coupling of the $f_0(980)$ to the $K\bar K$ system in isospin 0.

Further, we have used this amplitude to study the $f_0(980)\pi\pi $ system and coupled channels treating them like effective three-body systems.  This system is found to generate dynamically a $0^{++}$ resonance with mass $\sim$ 1773 MeV and width $\sim$ 100 MeV. The formation of this state in the $f_0(980)\pi\pi$ system occurs when both $f_0(980)\pi$ subsystems are found to generate the $\pi(1300)$. Such a structure makes that this $0^{++}$ resonance decays to $\pi\pi$, $\pi\pi\pi\pi$ and $\pi\pi K\bar K$, but not to $K\bar K$. These findings are not in agreement with the known scalar resonance in the 1700 MeV region, i.e., $f_0(1710)$, but are strikingly similar to the features of the recently claimed $f_0(1790)$ in the experimental data \cite{Bes1,Bugg,Bes2}. Thus, we relate our $0^{++}$
state with the $f_0(1790)$.

We have also studied the $\eta K\bar K$, $\eta\pi\pi$ systems in S-wave and no $\eta$ resonances or bound states are found in the energy region considered (1200-1900 MeV). It is worth mentioning here that we have studied several three hadron systems consisting of a meson or a baryon and a $K\bar{K}$ pair ($\phi K \bar{K}$~\cite{mko3}, $N K \bar{K}$~\cite{AJ2, JidoEnyo}, $K K \bar{K}$~\cite{AJ3} and now $\pi K \bar{K}$)
and in all these cases we found that the  $f_0 (980)$ configuration of the $K\bar{K}$ pair gives rise to a strong attraction in the three-body system which leads to the dynamical generation of a state. However, not enough attraction gets developed in the $\eta f_0(980)$ configuration of the $\eta K \bar{K}$ system to form
a bound state or resonance.

\section*{Acknowledgments}
The work of A.~M.~T.~is supported by  
the Grant-in-Aid for the Global COE Program ``The Next Generation of Physics, 
Spun from Universality and Emergence" from the Ministry of Education, Culture, 
Sports, Science and Technology (MEXT) of Japan.
This work is supported in part by
the Grant for Scientific Research (No.~22105507 and No.~22540275) from 
MEXT of Japan and the Grant-in-Aid for Scientific Research on Priority Areas titled Elucidation of New Hadrons with a Variety of FlavorsÓ (E01: 21105006) and the authors (K. P. K. and A. H.) acknowledge the same.
A part of this work was done in the Yukawa International Project for 
Quark-Hadron Sciences (YIPQS).

\end{document}